\documentstyle[preprint,aps]{revtex}
\tighten
\begin{document}
\draft
\preprint{
 \parbox[t]{50mm}{hep-ph/9408210 %bulletin Board\\
 \parbox[t]{50mm}{CHIBA-EP-84\\
 \parbox[t]{50mm}{DPNU-94-33\\ }}}}
\title{First-order Phase Transition in Three-dimensional QED
with Chern--Simons Term}
\author{K.--I. Kondo\cite{emkon}}
\address{Department of Physics, Chiba University, Chiba 263, Japan}
\author{P. Maris\cite{emmar}}
\address{Department of Physics, Nagoya University, Nagoya 464-01, Japan}
%\date{\today}
%
\maketitle
\begin{abstract}
We have studied the chiral phase transition in three-dimensional QED
in the presence of a Chern--Simons term for the gauge field. There
exists a phase where the chiral symmetry is broken dynamically and we
have determined the critical line for this symmetry breaking as a
function of the effective coupling and the strength of the additional
Chern--Simons term. In the presence of a Chern--Simons term, the
chiral phase transition turns out to be first order, in sharp contrast
to the phase transition in pure three-dimensional QED.
\end{abstract}

Dynamical symmetry breaking in three-dimensional quantum
electrodynamics (QED3) has attracted much attention over the last ten
years, both from a purely field theoretical point of view and because
of its applications to condensed matter physics in connection with
phenomena occurring in planar surfaces
\cite{qed3gen,nash,lat,craig,Fradkin91}. A natural extension of pure
QED3 is to add a Chern--Simons (CS) term for the gauge field
\cite{cs}, which breaks parity explicitly. Indeed the
statistics-changing CS term, together with the question whether or not
there is a dynamically generated fermion mass, plays a key role for
variants of QED3 to be effective theories for high-$T_c$
superconductivity and the fractional quantum Hall effect
\cite{Fradkin91}. Furthermore, QED3 also has implications for high
energy physics and physics of the early universe, since
three-dimensional models are the high temperature limit of the
corresponding four-dimensional theory.  Recently, it has been
suggested that the effective potential for high temperature QCD is
also related to CS gauge theories \cite{Nair93}.

The existence of the CS term leads to a novel feature in QED3, namely
a {\em first-order} chiral phase-transition, as we show in this
letter. The CS term generates a parity odd mass term for the fermions,
but in addition there might be a parity even mass, which breaks
chirality \cite{hopa,kon}. Dynamical chiral symmetry breaking, a
nonperturbative phenomenon, can be studied using the Dyson--Schwinger
(DS) equation for the full fermion propagator. Both a numerical study
of the full (truncated) DS equation and an analytical study of the
approximated equations show that there is a first-order phase
transition. This is quite remarkable and in sharp contrast to the
infinite order phase transition one finds using the same truncation
scheme in pure QED \cite{qed3gen,nash}.

The Lagrangian in Euclidean space is
\begin{eqnarray}
{\cal L} &=& \bar\psi(i \not{\!\partial} + e \not{\!\! A}
             - m_e - \tau m_o )\psi
        + {\textstyle{1\over4}} F_{\mu\nu}^2 \nonumber\\ &&
        + {\textstyle{1\over2}} i \theta \epsilon_{\mu\nu\rho}
                    A_\mu \partial_\nu A_\rho
        + {\cal L}_{\hbox{gauge fixing}} \,,
\end{eqnarray}
with the dimensionful parameter $\theta$ determining the relative
strength of the CS term. We use four-component spinors for the
fermions, and a four-dimensional representation for the
$\gamma$-matrices. The matrix $\tau$ is defined in such a way that the
term $m_o \bar\psi \tau \psi$ is odd under a parity transformation
\cite{kon,semwccw}. Also the CS term is odd under a parity
transformation, the other terms in the Lagrangian are invariant under
a parity transformation.

With such a representation we can define chirality similar as in
four-dimensional QED. Without an explicit mass $m_e$ for the fermions,
the Lagrangian is chirally symmetric, but the parity even mass $m_e$
breaks this symmetry. Note that the other mass, $m_o$, is chirally
invariant. Just as in pure QED, chiral symmetry can be broken
dynamically due to nonperturbative effects, which can be studied by
solving the DS equation for the fermion propagator with both explicit
masses $m_e$ and $m_o$ equal to zero.

The standard way to truncate the DS equation in QED3 is the $1/N$
expansion \cite{expansion}, where $N$ is the number of fermion
flavors. The coupling constant $e^2$ has the dimension of mass, and we
use the large $N$ limit in such a way that $e^2 \downarrow 0$ and the
product $N\,e^2$ remains fixed: $ N e^2 = 8 \alpha$ with $\alpha$
fixed. In this $1/N$ expansion the one-loop vacuum polarization has to
be taken into account, because this vacuum polarization is of order
one: using bare massless fermions, the transverse part of the vacuum
polarization is just $ \Pi^T(q) = - \alpha |q|$ \cite{qed3gen}. It is
easy to show that at order one there is no parity odd part of the
vacuum polarization.

The full vertex is replaced by the bare one, because that is the
leading order contribution in $1/N$. In order to be consistent with
the requirement that the vertex renormalization and the fermion wave
function renormalization are equal, we use a suitable nonlocal gauge
function \cite{nonloc}. In pure QED, one can construct a gauge in
which the wave function renormalization is exactly one. In the
presence of a CS term, this condition can only be satisfied up to
order $\theta/N$, but we are considering small $\theta$ only. The
proper choice for the gauge function is \cite{kon}
\begin{eqnarray}
a(q) & = & 2 \bigg( q^2 D^T(q) + \frac{2\alpha}{|q|}
  + \frac{4\alpha\theta}{q^2}
     \arctan{\frac{\theta |q|}{\alpha^2 + \alpha |q| + \theta^2}}
\nonumber \\ &&
  + \frac{\alpha^2 - \theta^2}{q^2}
  \ln{\frac{\alpha^2 + \theta^2}{(\alpha+|q|)^2 + \theta^2}} - 1\bigg) \,.
\end{eqnarray}
With this gauge, we also satisfy the Ward--Takahashi identity up to
corrections proportional to $\theta/N$ and to the dynamically
generated mass function, which are both negligible. Gauge covariance
can (in principle) be recovered by applying the Landau--Khalatnikov
transformation rules to the various Green's functions \cite{lk55}.

The inverse full fermion propagator can be written as
\begin{eqnarray}
 S^{-1}(p) &=& A_e(p) \not{\!p} + A_o(p)\tau \not{\!p}
                        - B_e(p) - B_o(p)\tau \,.
\end{eqnarray}
The functions $A(p)$ and $B(p)$ are scalar functions of the absolute
values of the momenta, and their bare values are $A_e = 1$, $A_o = 0$,
$B_e = m_e$, and $B_o = m_o$. We use the decomposition $A_\pm = A_e
\pm A_o$ and $B_\pm = B_e \pm B_o$, which leads, together with the
above truncation scheme, to the following two sets of coupled integral
equations
\begin{eqnarray}   \label{inteqA}
\lefteqn{ A_\pm(p) = 1 \pm \frac{8 \alpha}{N\, p^2} \int\!
           \frac{{\rm d^3}k}{(2\pi)^{\rm 3}}
       \frac{2 B_\pm(k) D^O(q) }{k^2 A_\pm^2(k) + B_\pm^2(k)}
     \, \frac{p\cdot q}{|q|} \,,} \\
\label{inteqB}
\lefteqn{ B_\pm(p) = \frac{8 \alpha}{N}
        \int\!\frac{{\rm d^3}k}{(2\pi)^{\rm 3}}
       \frac{1}{k^2 A_\pm^2(k) + B_\pm^2(k)} \times}  \nonumber\\ &&
           \Bigg(B_\pm(k)\left(2 D^T(q) + \frac{a(q)}{q^2} \right)
  \mp 2 A_\pm(k) D^O(q) \frac{k\cdot q}{|q|}\Bigg) \,,
\end{eqnarray}
where $D^T$ and $D^O$ are the transverse and the parity odd part of
the gauge boson propagator
\begin{eqnarray}
 D^T(q^2) &=& \frac{q^2 + \alpha |q|}
                        {(q^2 +\alpha |q|)^2 + \theta^2 q^2} \,,\\
 D^O(q^2) &=& \frac{- \theta |q|}
                        {(q^2 +\alpha |q|)^2 + \theta^2 q^2} \,,
\end{eqnarray}
and $q = k - p$. Note that the equations for $A_+$ and $B_+$ decouple
from the ones for $A_-$ and $B_-$. It is also important to observe
that once we have found a solution for $A_+$ and $B_+$, we {\em
automatically} have also a solution for $A_-$ and $B_-$: namely the
set $A_- = A_+$ and $B_- = - B_+$. That means that we can always
construct a chirally symmetric (but parity odd) solution, with $B_e =
0$.  The question of dynamical chiral symmetry breaking turns into the
question whether or not there exist {\em two} (or more) solutions of
the set of integral equations.

Without the CS term there is dynamical chiral symmetry breaking only
for $\lambda > \lambda_c = 3/16$ \cite{nash}, where we have defined
the effective coupling $\lambda = 8/(N \pi^2)$.  We expect a similar
situation in the presence of the CS term, at least if the parameter
$\theta$ is small. That means that for $\lambda < \lambda_c$ we only
have the chirally symmetric solution of the above equations, but for
$\lambda > \lambda_c$ we expect that there are (at least) two
solutions for both $B_+$ and $B_-$ possible, in such a way that there
is a nonzero solution for $B_e$. An essential difference from pure QED
is that in the presence of the CS term there is no trivial solution
$B_\pm(p) = 0$.  Due to the explicit breaking of parity, the fermions
always acquire a parity-odd mass term $B_o$, even if the explicit odd
mass term $m_o$ is zero.

Firstly, we solve the DS equation analytically after some further
approximations. Using $A(p) = 1 + {\cal O }(\theta)$ due to the
nonlocal gauge, we replace $A(p)$ by one, so we get an integral
equation for $B_\pm$ only, consisting of two terms, Eq.~(\ref{inteqB})
with $A_\pm(p)=1$. The first term is the same as in pure QED3, and the
essential region for this term is the infrared \cite{qed3gen}, $p,k <<
\alpha$. So we consider the integral for small momenta only, expand
the integration kernel in powers of $p$ and $k$, and introduce a
cutoff at $k = \alpha$. We also linearize the equation, by replacing
the denominator $k^2 + B_\pm^2(k)$ by $k^2 + B_\pm^2(0)$, which is
reliable as long as $B_\pm(p)$ is almost constant for small momenta.
In pure QED these approximations lead to almost the same result as the
full nonlinear integral equation \cite{thes,prep}.

The second term, proportional to $\theta$, can be calculated by
neglecting $B_\pm^2$ with respect to $k^2$ in the denominator, and
expanding the integrand in powers of $\min(p,q)/\max(p,q)$ and
$\theta$. Taking into account only the leading order terms gives in
the infrared region
\begin{eqnarray}
  I_\theta(p) & = & \mp \frac{8 \alpha}{N}
       \int\!\frac{{\rm d^3}k}{(2\pi)^{\rm 3}}
       \frac{2\, D^O(q)}{k^2 + B_\pm^2(k)}
            \frac{k\cdot q}{|q|} \nonumber \\
        & \simeq & \pm {\lambda\theta} + {\cal O}{(p)}
                         + {\cal O}{(\theta^3)} \,,
\end{eqnarray}
and in the ultraviolet region $p > \alpha $
\begin{eqnarray}
  I_\theta(p) & \simeq & \pm \frac{11\lambda\alpha\theta}{9 p}
            + {\cal O}{(1/p^2)} + {\cal O}{(\theta^3)} \,.
\end{eqnarray}
This means that in the ultraviolet region the CS term will dominate,
since without the CS term, $B(p)$ falls off much more rapidly in the
far ultraviolet. Higher order contributions in $\min(p,q)/\max(p,q)$
will slightly change this result, but not affect the general behavior
\cite{prep}.

Thus we have for $p < \alpha$ and to order $\theta$
\begin{eqnarray}
  B_\pm(p) & = & {\textstyle{4\over3}} \lambda \int_0^\alpha{\rm d}k\,
                \frac{k^2}{k^2 + M_\pm^2}
                       \frac{B_\pm(k)}{\max(p,k)}
                 \pm  {\lambda\theta} \,,
\end{eqnarray}
where we have defined $M_\pm = B_\pm(0)$. This integral equation can
easily be solved by converting it to a second-order differential
equation with boundary conditions.  The solution is
\begin{eqnarray}
  B_\pm(p) & = &
     M_\pm \, _2F_1(a_+,a_-,{\textstyle{3\over2}};-p^2/M_\pm^2) \,,
\end{eqnarray}
where $a_\pm = \frac{1}{4}(1 \pm i\sqrt{16\lambda/3 - 1})$. The
ultraviolet boundary condition leads to the condition
\begin{eqnarray}  \label{condition}
  M_\pm \, _2F_1(a_+,a_-,{\textstyle{1\over2}};-\alpha^2/M_\pm^2)
             & = & \pm \lambda\theta \,.
\end{eqnarray}

In order to determine the behavior of $M$ as a function of $\theta$,
we can plot the LHS of Eq.~(\ref{condition}) divided by $\lambda$ for
a given value of $\lambda$ as function of the mass parameter $M$, see
Fig.\ref{mvst} (for convenience we have set $\alpha = 1$ in our
figures, which just defines the energy scale).  From this figure we
can see that there are three solutions possible for $B_+$ and $B_-$ at
small values of $\theta$. A closer look at the region around the
origin would reveal that there exist more solutions for extremely
small values of $\theta$. In the absence of a CS term there are
infinitely many oscillating solutions \cite{qed3gen}, but it has been
shown that the vacuum corresponds to the nodeless one, with the
highest value for $|M|$. With the CS term, there is only a finite
number of oscillating solutions \cite{kon,prep}.

The chirally symmetric solution consist of the combination of $B_+(p)$
and $B_-(p)$ with $B_-(p) = - B_+(p)$, which can always be
constructed. A solution which breaks chiral symmetry can only be
constructed if there is a different solution $\tilde M_-$ of
Eq.~(\ref{condition}), corresponding to $\tilde B_-(p)$ which is {\em
not} equal to $-B_+(p)$. As we can see from Fig.\ref{mvst}, this is
only possible for $\theta < \theta_c(\lambda)$, beyond this critical
value there is only one solution possible for $B_+$ and $B_-$ which
automatically gives $B_e = 0$ and $B_o = B_+ = -B_-$. At the critical
value $\tilde M_\pm$ does not become zero, nor does $B_e(0) = (B_+(0)
+ \tilde B_-(0))/2$, which can be regarded as the order parameter of
the chiral phase transition. This clearly signals a first-order phase
transition, in sharp contrast to the pure QED case.

We can also plot $M_\pm$ versus $\lambda$ for a fixed value of
$\theta$, see Fig.~\ref{mvsl}.  Here we see that if we increase
$\lambda$ for fixed $\theta$, the chiral symmetry breaking solutions
appears if $\lambda$ exceeds some critical value $\lambda_c(\theta)$,
which increases rapidly as a function of $\theta$. This figure shows
that the chiral phase transition is first order in this direction as
well: increasing $\lambda$ beyond $\lambda_c(\theta)$ gives rise to
the second (and third) solution, but at the phase transition neither
$\tilde M_\pm$ nor $B_e(0)$ become zero. In this figure we can also
see that in the limit $\theta \rightarrow 0$ both $M_\pm$ and $\tilde
M_\pm$ go towards the nontrivial solution $m$ of pure QED, and the
critical value $\lambda_c$ goes towards $\lambda_c (\theta = 0) =
3/16$.

The critical parameters $\lambda_c$ and $\theta_c$ can be calculated
directly from Eq.~\ref{condition}. In Fig.~\ref{critlt} we have shown
the critical line in the $(\lambda,\theta)$-plane. For small values of
$M$ we can make an expansion in order to get an explicit expression
for $\theta_c(\lambda)$.  To leading order in $\sqrt{16\lambda/3 -
1}$, this gives
\begin{eqnarray}
\theta & \sim & \exp{(-3\pi/\sqrt{16\lambda/3 - 1})} \,.
\end{eqnarray}

Secondly, we have solved numerically the set of coupled integral
equations for $A$ and $B$, Eqs.~\ref{inteqA} and \ref{inteqB}, {\it
without any further approximations}, and these numerical results are
qualitatively in good agreement with our analytical results. First we
checked our assumption that $A_\pm(p)$ is close to one, and it turns
out that the deviation is indeed negligible for small values of
$\theta$ \cite{prep}.  Furthermore, we have found numerically the
following solutions for $B_+$, using the notation $m(p)$ for the
solution with $\theta = 0$:
\begin{enumerate}
\item
$B_+(p) = {\cal O}(m(p)) > 0$  for $\lambda > \lambda_c(\theta=0)$,\\
$B_+(p) = {\cal O}(\theta)$ for $\lambda < \lambda_c(\theta=0)$;
\item
$\tilde B_+(0) = {\cal O}(-m(0)) < 0$.
\end{enumerate}

The first solution exists for all values of both $\lambda$ and
$\theta$, whereas the second one exists only for values of $\lambda >
\lambda_c(\theta)$ and $\theta < \theta_c(\lambda)$. Note that this
second solution behaves like $\theta / p$ in the ultraviolet, so it
has some node at a particular value of $p$. The existence of this
second solution allows for a nonzero chiral symmetry breaking solution
$B_e(p) = (B_+(p) + \tilde B_-(p))/2$.

Numerically, it is extremely difficult to establish a first-order
phase transition and to determine the critical values $\lambda_c$ and
$\theta_c$. However, our numerical results all support our analytical
results, and indicate strongly that the chiral phase transition is
indeed first order. In Fig.~\ref{bvsl} we have shown the behavior of
$B_+(0)$ and $\tilde B_-(0)$ at fixed $\theta$ as a function of
$\lambda$. We can see that the behavior is the same as in
Fig.~\ref{mvsl}: increasing $\lambda$ at fixed $\theta$ leads to a
second solution $\tilde B_\pm$ beyond some critical value $\lambda_c >
\lambda_c(\theta=0)$. Close to the critical value $\tilde B_\pm(0)$
does not go to zero, nor does $B_e(0)$, signaling a first-order chiral
phase transition. Also the behavior for increasing $\theta$ at fixed
$\lambda$ is qualitatively the same as our analytical result.

In conclusion, both the numerical and the analytical results show that
there is a first-order chiral phase transition in QED3 with explicit
CS term. This result is very remarkable, given the well-known infinite
order phase transition (Miransky-scaling) \cite{qed3gen} in the
absence of the CS term. Also the other known chiral phase transitions
in four-dimensional gauge theories are of second (or higher) order.
This first-order phase transition is a new and interesting phenomenon,
and it might lead to new insights into chiral phase transitions in
general. In particular, the connection between the CS term and the
first-order phase transition should be studied in more detail.

This result should also be contrasted with some previous results in
analyzing this model \cite{hopa}, indicating just a minor quantitative
effect on the critical coupling and scaling behavior due to the CS
term. Both our numerical and analytical results reveal that the
presence of an explicit CS term changes the nature of the chiral phase
transition drastically.

This work was initiated during a visit of one of the authors (P. M.)
to Chiba University, and he would like to thank the members of the
Graduate School of Science and Technology for their hospitality during
that stay. We would like to thank K.~Yamawaki, Yoonbai Kim, and
T.~Ebihara for stimulating discussions. This work has been financially
supported by the JSPS (P. M. being a JSPS fellow under number 94146).

%%%%%%%%%%%%%%%%%%%%%%%%%  References  %%%%%%%%%%%%%%%%%%%%%%%

%%%%%%%%%%%%%%%%%%%%%%%%% Figures (captions) %%%%%%%%%%%%%%%%%%%%

\begin{figure}
\caption{
$\theta$ as function of $M$ for some different values of $\lambda$.
The upper half of the plane corresponds to solutions for $B_+$, the
lower half for $B_-$. Note that both the upper and the lower half
correspond to positive values of $\theta$.}
\label{mvst}
\end{figure}

\begin{figure}
\caption{
$|M_\pm|$ as function of $\lambda$ for some different values of $\theta$.}
\label{mvsl}
\end{figure}

\begin{figure}
\caption{
The critical line for the chiral phase transition in the $(\lambda,
\theta)$-plane.}
\label{critlt}
\end{figure}

\begin{figure}
\caption{
The infrared values $B_+(0)$ and $\tilde B_-(0)$ as function of
$\lambda$ for some different values of $\theta$.}
\label{bvsl}
\end{figure}

\end{document}